\documentclass[11pt,a4paper]{article}
\usepackage{jheppub}
\usepackage{cases}
\usepackage{graphicx}
\usepackage{dcolumn,color}
\usepackage{bm}
\usepackage{epstopdf}
\usepackage{epsfig}
\usepackage{color,framed}
\usepackage{amsmath,amsthm,amssymb} 

\usepackage[table]{xcolor}

\title{$K$-field kinks: stability, exact solutions and new features}

\author[a,b]{Yuan Zhong,}
\author[a,1]{Yu-Xiao Liu,\note{The corresponding author.}}

\affiliation[a]{Institute of Theoretical Physics, Lanzhou University,\\
 Lanzhou 730000, People' s Republic of China}
\affiliation[b]{IFAE, Universitat Aut\`onoma de Barcelona, 08193 Bellaterra, Barcelona, Spain}

\emailAdd{yzhong@ifae.es, liuyx@lzu.edu.cn}

\abstract{We study a class of noncanonical real scalar field models in $(1+1)$-dimensional flat space-time. We first derive the general criterion for the classical linear stability of an arbitrary static soliton solution of these models. Then we construct first-order formalisms for some typical models and derive the corresponding kink solutions. The linear structures of these solutions are also qualitatively analyzed and compared with the canonical kink solutions.}

\keywords{Solitons Monopoles and Instantons, Integrable Field Theories}

\begin{document}
\maketitle


\section{Introduction}
In field models with degenerate vacua, the classical fields might have nontrivial static solutions called solitons, which interpolate between vacua. Solitons have been studied extensively in the past decades in both condensed matter physics~\cite{BishopSchneider1978} and particle physics~\cite{RebbiSoliani1984}. The simplest solitons are kinks, which are  solutions of two-dimensional models with only a single real scalar field. In higher-dimensional space-time, a static soliton solution in canonical scalar field models was forbidden by the Derrick theorem~\cite{Derrick1964}. However, high-dimensional solitons do exist if the scalar field has noncanonical kinetic terms~\cite{Diaz-Alonso1983,Babichev2006,AvelinoBazeiaMenezes2011}.
Noncanonical scalar fields (also known as $K$-fields) have been repeatedly studied recently in cosmology~\cite{Armendariz-PiconDamourMukhanov1999,GarrigaMukhanov1999,Armendariz-PiconMukhanovSteinhardt2001}, string theories~\cite{Sen2002,Sen2003,Sen2005}, brane world models~\cite{AdamGrandiSanchez-GuillenWereszczynski2008,BazeiaGomesLosanoMenezes2009,LiuZhongYang2010,ZhongLiu2013,ZhongLiuZhao2014a}, and massive gravity theories~\cite{Hinterbichler2012,Rham2014}.

In the present paper, we focus on kink solutions in $K$-field models. Lots of interesting $K$-field models and corresponding kink solutions have been studied in
literature~\cite{BazeiaLosanoMenezesOliveira2007,
AdamSanchez-GuillenWereszczynski2007,BazeiaLosanoMenezes2008,SantosRubiera-Garcia2011,AlmeidaBazeiaLosanoMenezes2013,BernardiniBertolami2013,Rubiera-GarciaSantos2014}.
In ref.~\cite{BazeiaLosanoMenezesOliveira2007}, the authors first proposed the first-order formalisms for two types of $K$-field models and analyzed the linear stability of their solutions. These first-order formalisms help one to rewrite the original dynamical equations into some first-order ones that are easier to solve analytically. Later, the authors extended the first-order formalism method to more complex $K$-field systems~\cite{BazeiaLosanoMenezes2008,AlmeidaBazeiaLosanoMenezes2013}.

However, the study on $K$-field kinks is still incomplete.
First, the previous discussions on the linear stability of the kink solutions are not general and elegant enough for analytical analysis. Second, the first-order formalisms proposed in refs.~\cite{BazeiaLosanoMenezesOliveira2007,BazeiaLosanoMenezes2008,AlmeidaBazeiaLosanoMenezes2013} are too complex to obtain exact soliton solutions. In fact, in refs.~\cite{BazeiaLosanoMenezes2008,AlmeidaBazeiaLosanoMenezes2013} the authors only solved some $K$-field models perturbatively. Also, it is still unclear how the properties (such as the linear structure) of the noncanonical kinks would differ from those of the canonical ones.

In this paper, we report our study on the above issues. In the next section, we first derive the linear perturbation equation for a large class of noncanonical models. We find that linear perturbation around an arbitrary static solution satisfies a Schr\"odinger-like equation. We show that the Hamiltonian of the perturbation can be factorized, so that supersymmetric quantum mechanics can be applied to analyze the spectrum of the linear perturbation. Then, in secion \ref{sec3} we propose alternative first-order formalisms for two typical types of noncanonical models. Exactly solvable models can be easily constructed by using our formalisms. To illustrate this, we consider two explicit models, and give the corresponding kink solutions, which we call the generalized Sine-Gordon kink and the generalized $Z_2$ kink.
In section \ref{sec4} we compare the linear structure of one of our solutions with the canonical Sine-Gordon kink and $Z_2$ kink. Finally, we give a short summary on our results.

\section{Two-dimensional $K$-field model and the stability issue}
\label{sec2}
Most of the currently interested noncanonical scalar field models are described by the following action:
\begin{eqnarray}
\label{action}
S=\int dx dt \mathcal{L}(\phi,X),
\end{eqnarray}
where $X\equiv-\frac12\eta^{\mu\nu}\partial_\mu\phi\partial_\nu\phi$ (with $\eta^{\mu\nu}=\textrm{diag}(-1,1)$) represents the kinetic term of the scalar field $\phi$.
The equation of motion reads
\begin{eqnarray}
\label{dynamical}
\frac{{\partial {\cal L}}}{{\partial \phi }} + {\partial _\mu }\left(\frac{{\partial {\cal L}}}{{\partial X}}{\partial ^\mu }\phi\right ) = 0.
\end{eqnarray}
In this paper, we only study static soliton solutions, which means that $\phi=\phi(x)$ is independent of $t$, and the dynamical equation reduces to
\begin{eqnarray}
\label{EoM}
-\mathcal{L}_\phi=(\mathcal{L}_X \phi')'.
\end{eqnarray}
Here, we have defined $\mathcal{L}_\phi\equiv\frac{\partial \mathcal{L}}{\partial \phi}$ and $\mathcal{L}_X\equiv\frac{\partial \mathcal{L}}{\partial X}$. A prime always represents the derivative with respect to $x$.

The energy density (the Hamiltonian density) for a static solution is simply
\begin{eqnarray}
\rho=-\mathcal{L}.
\end{eqnarray}
Regardless of the exact mathematical definition of solitons, in this paper, we refer to a soliton as a solution of eq.~\eqref{EoM} that has a spatially localized energy density $\rho$~\cite{Rajaraman1982}.

Before we launch our study for exactly solvable models, let us first analyze the stability of an arbitrary candidate solution $\bar{\phi}(x)$ against a small oscillation $\delta\phi(x,t)$. To do this, we need to derive the linear order equation of $\delta\phi(x,t)$. Since our system is of second order, we are expected to obtain also a second-order equation for $\delta\phi(x,t)$. This equation usually can be recast into a Schr\"odinger-like equation. A solution is stable if and only if none of the energy eigenvalues of the corresponding Schr\"odinger-like equation are negative~\cite{Coleman1985}. Although the stability issue of $K$-field models was discussed in literature, none of them offers a general and mathematically simple criterion for stable $K$-defects. In what follows, we derive the most general stability condition for a kink solution of our model.

There are two typical methods to linearize a system. One can either expand the action into second-order of $\delta\phi$, or directly linearize the equation of motion. Naively, both approaches should be equivalent and lead to a same linear perturbation equation. However, in some cases, these two equations do not coincide. Some examples can be found in thick brane world models~\cite{Giovannini2001a,ZhongLiu2013}. Therefore, it is necessary to linearize our system with both methods.

\subsection{The quadratic action}
The second-order perturbation of the Lagrangian density $\mathcal{L}$ is
\begin{eqnarray}
  {\delta ^{(2)}}\mathcal{L}
 =
   {\mathcal{L}_X}{\delta ^{(2)}}X
  + \frac{1}{2}{\mathcal{L}_{\phi \phi }}{({\delta }\phi )^2}+ {\mathcal{L}_{\phi X}}{\delta}\phi {\delta ^{(1)}}X
  + \frac{1}{2}{\mathcal{L}_{XX}}{({\delta ^{(1)}}X)^2}.
\end{eqnarray}
For a static background solution, the first- and second-order perturbations of $X$ are
\begin{eqnarray}
\label{delta1X}
  {\delta ^{(1)}}X&=&-\delta \phi '\phi ', \\
\label{delta2X}
  {\delta ^{(2)}}X&=&- \frac{1}{2}({\partial^\mu}\delta \phi)( {\partial _\mu}\delta \phi).
\end{eqnarray}
Thus,
\begin{eqnarray}
\label{Lm}
&&{\delta ^{(2)}}{\cal L} = \frac{1}{2}\{ {{\cal L}_{\phi \phi }}{(\delta \phi )^2} + {{\cal L}_{XX}}{(\phi ')^2}{(\delta \phi ')^2}
- {{\cal L}_X}{\partial ^\mu}\delta \phi {\partial _\mu}\delta \phi  - 2{{\cal L}_{\phi X}}\phi '\delta \phi \delta \phi '\}.
\end{eqnarray}
This result consists with the one of ref.~\cite{BazeiaLosanoMenezesOliveira2007}.

From ${\cal L}_X' = {{\cal L}_{X\phi }}\phi ' + {{\cal L}_{XX}}X'$, we obtain
\begin{eqnarray}
\label{Lxp}
{{\cal L}_{X\phi }}=\frac{1}{\phi '}\left({\cal L}_X'- {{\cal L}_{XX}}X'\right).
\end{eqnarray}
Taking the derivative of the equation of motion, we get
\begin{eqnarray}
\mathcal{L}_\phi ' = \mathcal{L}_{\phi \phi }\phi ' + \mathcal{L}_{X\phi }X' =  - (\mathcal { L}_X\phi ')''.
\end{eqnarray}
Plugging eq.~\eqref{Lxp} into the above equation, we obtain the expression of $\mathcal{L}_{\phi\phi}$:
\begin{eqnarray}
\label{Lpp}
{{\cal L}_{\phi \phi }} =  - \frac{{({{\cal L}_X}\phi ')''}}{{\phi '}}
+\mathcal{L}_X'\frac{{\phi ''}}{{\phi '}} + {{\cal L}_{XX}}{{\phi ''}^2},
\end{eqnarray}
where, we have used $X'=-\phi'\phi''$.

Eliminating $\mathcal{L}_{X\phi}$ and $\mathcal{L}_{\phi\phi}$ by using eqs.~\eqref{Lxp} and \eqref{Lpp},  and defining $\mathcal{G}\equiv \delta \phi\sqrt{\mathcal{L}_X}$, we obtain
\begin{equation}
 {\delta ^{(2)}}\mathcal{L} = \frac{1}2\left\{
   -\mathcal{G}\partial_t^2\mathcal{G} +V(x)\mathcal{G}^2
+\gamma\mathcal{G}\mathcal{G}''\right\},
\end{equation}
where
\begin{equation}
V(x)=-\gamma \frac{z''}{z}-\frac{z'}{z}\gamma '-\frac{1}{2}\gamma '',
\end{equation}
and
\begin{equation}
\label{zf}
z=\phi '\mathcal{L}_X^{1/2},\quad
\gamma=1+2\frac{\mathcal{L}_{XX} X}{\mathcal{L}_X}.
\end{equation}

In the case $\gamma>0$, we can use the Regge-Wheeler ``tortoise'' coordinate $x^{\ast}$
\begin{equation}
\label{RWcoord}
\frac{dx^{\ast}}{dx}\equiv\gamma^{-1/2}
\end{equation} to rewrite the quadratic action as
\begin{eqnarray}
{\delta ^{(2)}}{S_\mathcal{G}} &= &\frac{1}{2}\int dtdx^{\ast}{\sqrt{\gamma}}\times
 \left\{
   -\mathcal{G}\partial_t^2\mathcal{G} +V_{\text{eff}}(x^{\ast})\mathcal{G}^2
+ \mathcal{G} \ddot{\mathcal{G}}\right\},
\end{eqnarray}
with
\begin{eqnarray}
V_{\text{eff}}(x^{\ast})&\equiv & V(x^{\ast})+\frac{1}{4\sqrt\gamma}\frac{d}{dx^{\ast}}\left(\frac{ \dot\gamma}{ \sqrt{\gamma }}\right).
\end{eqnarray}
Here, a over dot represents the derivative with respect to $x^\ast$.

Obviously, the normal modes of the quadratic action is
\begin{equation}
\hat{\mathcal{G}}=\frac{1}{\sqrt{2}}\gamma^{1/4}\mathcal{G}.
\end{equation}
In terms of $\hat{\mathcal{G}}$, the quadratic action reads
\begin{equation}
\label{G2}
 {\delta ^{(2)}}{S_{\hat{\mathcal{G}}}} = \int dtdx^{\ast}\hat{\mathcal{G}}
 \left\{-\partial_t^2\hat{\mathcal{G}}+ \ddot{\hat{\mathcal{G}}} -\frac{\ddot{\theta}}{\theta}\hat{\mathcal{G}}
\right\},
\end{equation}
where
\begin{equation}
\theta \equiv\gamma^{1/4}z .
\end{equation}

From the quadratic action of $\hat{\mathcal{G}}$, we know that for
\begin{eqnarray}
\label{criterion}
\mathcal{L}_X>0,\quad \gamma>0,
\end{eqnarray}
the linear perturbation satisfies a Schr\"odinger-like equation
\begin{equation}
\label{schroScalar}
-\ddot{\hat{\mathcal{G}}} +\frac{\ddot{\theta}}{\theta}\hat{\mathcal{G}}
=-\partial_t^2\hat{\mathcal{G}}.
\end{equation}

\subsection{The linear perturbation equation}
The linear perturbation equation can also be derived directly from the equation of motion. Linearizing eq.~\eqref{dynamical} we obtain
\begin{eqnarray}
&&{{\cal L}_X}{\partial _\mu }{\partial ^\mu }\delta \phi +\delta {{\cal L}_\phi } + \delta {\cal L}_X'\phi '
+{\cal L}_X'\delta \phi '
+ \delta {{\cal L}_X}\phi ''= 0.
\end{eqnarray}
Using the following identities:
\begin{eqnarray}
\delta {{\cal L}_\phi } &=&  - {{\cal L}_{X\phi }}\phi '\delta \phi ' + {{\cal L}_{\phi \phi }}\delta \phi, \\
\delta {{\cal L}_X} &=& {{\cal L}_{X\phi }}\delta \phi  - {{\cal L}_{XX}}\phi '\delta \phi',
\end{eqnarray}
as well as eqs.~\eqref{Lxp} and~\eqref{Lpp}, we finally obtain the following equation
\begin{eqnarray}
\label{26}
&&\gamma \left( {\left( \partial _x\ln(\gamma {{\cal L}_X}) \right)\frac{{X'}}{{ - 2X}} - \frac{{X''}}{{2X}} + \frac{1}{4}{{\left( {\frac{{X'}}{X}} \right)}^2}} \right)\delta \phi\nonumber\\
&+&\gamma \left( {\partial_x}{\ln(\gamma {{\cal L}_X})} \right)\delta \phi '
+ \gamma \delta \phi '' - \partial _t^2\delta \phi    = 0.
\end{eqnarray}
Defining
\begin{eqnarray}
\psi=\delta\phi \sqrt{\gamma\mathcal{L}_X},
\end{eqnarray}
eq. \eqref{26} reduces to
\begin{eqnarray}
\gamma  \psi ''-\gamma \frac{\left(z\sqrt{\gamma }\right)''}{z\sqrt{\gamma }}\psi -\partial_t^2\psi=0.
\end{eqnarray}

To proceed, we introduce the $x^\ast$ coordinates, and redefine $\psi=\gamma^{1/4}\varphi$, in the end, we obtain a Schr\"odinger-like equation
\begin{equation}
-\ddot{\varphi} +\frac{\ddot{\theta}}{\theta}\varphi
=-\partial_t^2\varphi.
\end{equation}
This is the same equation we obtained from the quadratic action in eq.~\eqref{G2} (by definition $\varphi=\sqrt{2}\hat{\mathcal{G}}$). It is interesting to note that when gravity is tuned on, the equations obtained in this two approaches (namely, linear equation and quadratic action) are different. But here, in the case without gravity, we find that the equations are equivalent.

\subsection{Supersymmetric quantum mechanics and the energy spectrum}
In this subsection, we analyze some general features of the model described by eq.~\eqref{action}. First, we make the following decomposition
\begin{equation}
\hat{\mathcal{G}}=\sum_{n=0}^{\infty} f_n(x^\ast)e^{i\omega_n t}.
\end{equation}
Then we find that the equation for $f_n(x^\ast)$ is
\begin{equation}
H f_n=-\ddot{f}_n+\mathcal{V} f_n=\omega_n^2 f_n,
\end{equation}
where $H=-\frac{d^2}{dx^{\ast 2}}+\mathcal{V}$ is the linear perturbation Hamiltonian, and $\mathcal{V}=\ddot{\theta}/{\theta}$ is the effective potential.
An important property of this Hamiltonian is that it can be factored as
\begin{equation}
\label{hatG}
H=\mathcal{A}\mathcal{A}^\dagger,
\end{equation} where
\begin{equation}
\label{Adagger}
\mathcal{A}=\frac{d}{dx^{\ast}}+\frac{\dot{\theta}}{\theta},\quad
\mathcal{A}^\dagger=-\frac{d}{dx^{\ast}}+\frac{\dot{\theta}}{\theta}.
\end{equation}
Systems with factorable Hamiltonians have been extensively studied in supersymmetic quantum mechanics~\cite{CooperKhareSukhatme1995}, and they have two important properties:
\begin{enumerate}
  \item $\omega_n$ are semipositive definite, namely, $\omega_n\geq 0$. The zero mode ($\omega_0=0$) of $H$ reads
\begin{eqnarray}
f_0=c  \theta(x^{\ast}),
\end{eqnarray} where $c$ is the normalization constant.
  \item One can construct a partner Hamiltonian
\begin{eqnarray}
H_{-}&=&\mathcal{A}^\dagger \mathcal{A}
=-\frac{d^2}{dx^{\ast 2}}+\mathcal{V}_{-},
\end{eqnarray}
where $\mathcal{V}_{-}\equiv\theta\left(\theta^{-1}\right)^{..}$. Except the zero mode $f_0$, $H_{-}$ and $H$ share the same spectrum.
\end{enumerate}
The first property tells us that any background solution that satisfies the inequalities \eqref{criterion} is always stable against small linear perturbation. While the second one offers us an alternative way to analyze the mass spectrum of the linear perturbation. As we will show in section \ref{sec4}, $H_{-}$ enables us to discern the pattern of the eigenstates of $H$ easily.

\section{First-order formalisms and kink solutions}
\label{sec3}
With the stability criterions, now we can establish solvable noncanonical models that support stable kink solutions.

For canonical scalar field $\mathcal{L}=X-V$, analytically solvable models can be constructed via the so called first-order formalism (also known as the superpotential method ~\cite{Vachaspati2006}). In this formalism, both the scalar $\phi$ and the scalar potential $V$ are rewritten in terms of the so called superpotential $W(\phi)$, which is an arbitrary function of $\phi$:
\begin{eqnarray}
\label{canoPhi}
\phi'&=&W_\phi,\\
V&=&\frac12 W_\phi^2.
\end{eqnarray}
Once the first-order formalism is established, one can easily construct a solvable model by simply specify the form of $W$. For some carefully chosen $W$, the corresponding model supports kink-like soliton solutions.

The first-order formalism was also extended to some $K$-field models~\cite{BazeiaLosanoMenezes2008,BazeiaGomesLosanoMenezes2009,AlmeidaBazeiaLosanoMenezes2013}.
In these papers, the authors generalized eq.~\eqref{canoPhi} to
\begin{eqnarray}
\label{fomalism1}
\mathcal{L}_X\phi'=W_\phi.
\end{eqnarray}
Applying the above equation to a model with $\mathcal{L}=X-\alpha X^2-V$, the authors obtained
\begin{eqnarray}
\phi'+\alpha\phi'^3=W_\phi.
\end{eqnarray}
The above equation leads to a complex relation between $\phi'$ and $W_\phi$. To proceed, the authors assumed $\alpha$ to be a small parameter, and treated the noncanonical model perturbatively.

In this section, we argue that many exactly solvable models can be constructed if one starts with\footnote{Note that we use $W$ rather than $W_\phi$ for simplicity, once we know $W$ it is easy to get $W_\phi$ and vise versa.}
\begin{eqnarray}
\label{formalism2}
\phi'=W.
\end{eqnarray}
To see this, let us consider two types of models:
\begin{itemize}
  \item Type I: $\mathcal{L}=F(X)-V(\phi)$;
  \item Type II: $\mathcal{L}=F(X)U(\phi)$.
\end{itemize}

\subsection{Type I models}
For the type I models, the equation of motion reads
\begin{eqnarray}
V_\phi=(F_X \phi')',
\end{eqnarray}
or
\begin{eqnarray}
d V&=&\phi'd(F_X\phi').
\end{eqnarray}
Using eq.~\eqref{formalism2}, we obtain
\begin{eqnarray}
\label{v1}
\partial_W V&=&W \partial_W(F_X W).
\end{eqnarray}
Since $F_X$ is a function of only $W^2$, the expression for $V(W)$ can be obtained by merely an integration.

For example, we consider the $X^2$ model
\begin{eqnarray}
\label{model1}
\mathcal {L}=X-\alpha X^2-V.
\end{eqnarray}
Obviously, for $\alpha\geq 0$, the stable criterion \eqref{criterion} is always satisfied, and any solution of this model will be linearly stable.
From eq.~\eqref{v1}, we obtain
\begin{eqnarray}
\label{V}
V=\frac12 W^2+\frac34 \alpha W^4+V_0.
\end{eqnarray}
For simplicity, let us take $V_0=0$.
Then, the energy density reads
\begin{eqnarray}
\label{rhoX2}
\rho=W^2+\alpha W^4.
\end{eqnarray}
Equations \eqref{formalism2} and \eqref{V} constitute the first-order formalism of the $X^2$ model.

\subsection{Type II models}
For the type II models, the equation of motion is
\begin{eqnarray}
-FU_\phi=\frac d{dx}(U F_X \phi').
\end{eqnarray}
Using the superpotential, one obtains
\begin{eqnarray}
\label{Type2}
\partial_W \ln U=-W\frac{F_X+W \partial_W F_X}{F+W^2F_X}.
\end{eqnarray}
After an integration, we can easily obtain the expression of $U(W)$. Equations \eqref{Type2} and \eqref{formalism2} constitute the first-order formalism of the type II models. Given a suitable $W$, one would get an analytically solvable model.

One of the typical type II models is the scalar Born-Infeld model, where
\begin{eqnarray}
\label{BI}
\mathcal{L}=-U(\phi)\sqrt{1-X}+\varepsilon_0.
\end{eqnarray}
Obviously, the stability criterions are satisfied if $U(\phi)>0$.
Applying eq.~\eqref{Type2}, we obtain
\begin{eqnarray}
U=U_0 \sqrt{2+W^2}, \quad U_0>0.
\end{eqnarray}
Taking $U_0=\sqrt{2}$ and $\varepsilon_0=2$, we obtain the expression of the energy density
\begin{eqnarray}
\label{rhoTachyon}
\rho=W^2.
\end{eqnarray}

Comparing eq.~\eqref{rhoTachyon} with eq.~\eqref{rhoX2}, one finds that the tachyon model and the canonical model (with $\alpha=0$) can have not only the same field configuration (by solving eq.~\eqref{formalism2}), but also the same energy density (by taking $V_0=0$). Models with this feature are called the twinlike models, which were first observed in~\cite{AndrewsLewandowskiTroddenWesley2010}, and later developed in~\cite{BazeiaMenezes2011,AdamQueiruga2011,BazeiaDantasGomesLosanoMenezes2011,AdamQueiruga2012,BazeiaDantas2012,BazeiaLobaoMenezes2012,GomesMenezesNobregaSimas2013}.
Note that, although twinlike models are indistinguishable in the background level, usually their linear perturbation structures are different (see ref.~\cite{AndrewsLewandowskiTroddenWesley2010}). In fact from eq.~\eqref{schroScalar} we know that twinlike models have same linear perturbation structures if and only if they have a same $\theta$. This is a strong constraint that most twinlike models do not satisfy. So, linear perturbation can be used to distinguish most twinlike models. For instance, the Born-Infeld model described by eq.~\eqref{BI} is a distinguishable twin model of the canonical model~\cite{AndrewsLewandowskiTroddenWesley2010}. However, there are some interesting cases where twinlike models even possess the same linear perturbation structures~\cite{BazeiaMenezes2011,AdamQueiruga2012}.

\subsection{Kink solutions}
So far, we have established the first-order formalisms for two classes of typical $K$-field models. Now we are ready to construct analytically solvable models by simply giving some suitable forms of $W$. For instance, the following two types of superpotentials lead to two different kink solutions:
\begin{eqnarray}
\label{W}
W_1&=&k v \cos \left(\frac\phi{v}\right),\\
W_2&=&kv\left(1-\left(\frac{\phi }{v}\right)^2\right),
\end{eqnarray}
where both $k$ and $v$ are positive constants. In the canonical case, $W_1$ leads to the Sine-Gordon model, and $W_2$ corresponds to the $Z_2$ symmetric $\phi^4$ model. The scalar configurations corresponding to $W_1$ and $W_2$ are
\begin{eqnarray}
\label{phi1}
\phi_1&=&v \arcsin(\tanh(k x)),\\
\label{phi2}
\phi_2&=&v\tanh \left(k x\right),
\end{eqnarray} respectively. Obviously, $v$ represents the vacuum expectation value of $\phi(x)$, and $1/k$ the thickness of the soliton.
\begin{figure*}
\begin{center}
\includegraphics[width=0.48\textwidth]{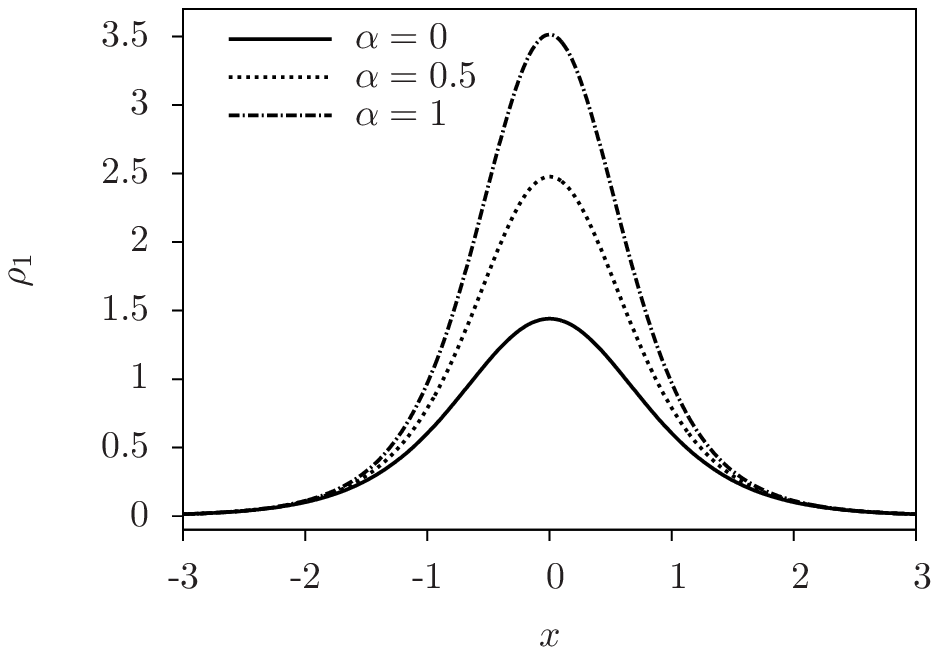}
\includegraphics[width=0.48\textwidth]{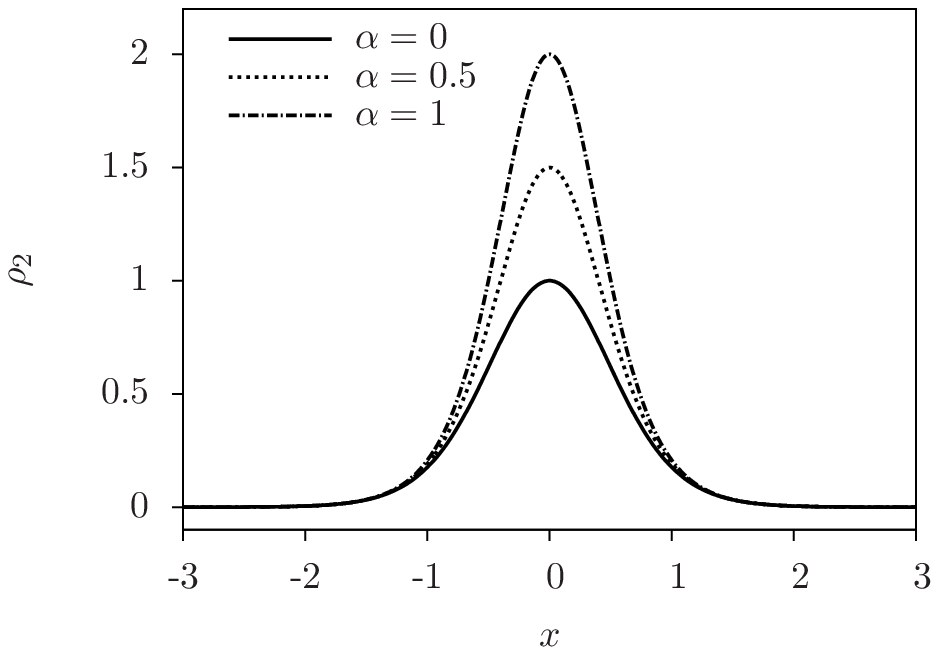}
\end{center}
\caption{Energy densities of the solutions of the $X^2$ model. The parameters are set to $k=1$ and $v=1.2$.}
\label{figureEnergyDensity}
\end{figure*}
Plugging the expressions of $W_i(\phi)$ and $\phi_i(x)$ ($i=1,2$) into $V(W)$ and $\rho(W)$, one can easily obtain analytical forms of $V_i(\phi)$ and $\rho_i(x)$. This is a trivial work so we will not do it here. Instead, we plot the energy densities of the solutions of the $X^2$ model in figure~\ref{figureEnergyDensity}, form which we see that the energy densities are indeed localized around $x=0$. Therefore, the generalized kink solutions are soliton solutions. The Born-Infeld model is equivalent to the canonical model at the background level, so the energy density is the solid lines in figure~\ref{figureEnergyDensity}.

\section{Linear structure of $X^2$ model}
\label{sec4}
The linear structures of both the standard Sine-Gordon model and $Z_2$ symmetric $\phi^4$ model are well-known in literature~\cite{Vachaspati2006}:
\begin{itemize}
  \item The spectrum of the perturbation Hamiltonian $H$ in the Sine-Gordon model is consist of the zero mode and a continuum of states which are separated from the zero mode by a mass gap.
  \item In the $Z_2$ model, however, $H$ has one more discrete bound state in addition to the zero mode and the continuum of states.
\end{itemize}
In this section, we discuss the possible influences from the $X^2$ term on the above well-known conclusions. Note that, for all the type I models the effective Schr\"odinger potential is determined only by the form of $F(X)$. So the $X^2$ model is a good toy model for us to analyze how noncanonical kinetic terms would affect the properties of the canonical models. For simplicity, we refer to the soliton solution generated by $W_1$ as the generalized Sine-Gordon kink, and the one generated by $W_2$ as the generalized $Z_2$ kink.

\subsection{The generalized Sine-Gordon kink}
\begin{figure*}
\begin{center}
\includegraphics[width=0.48\textwidth]{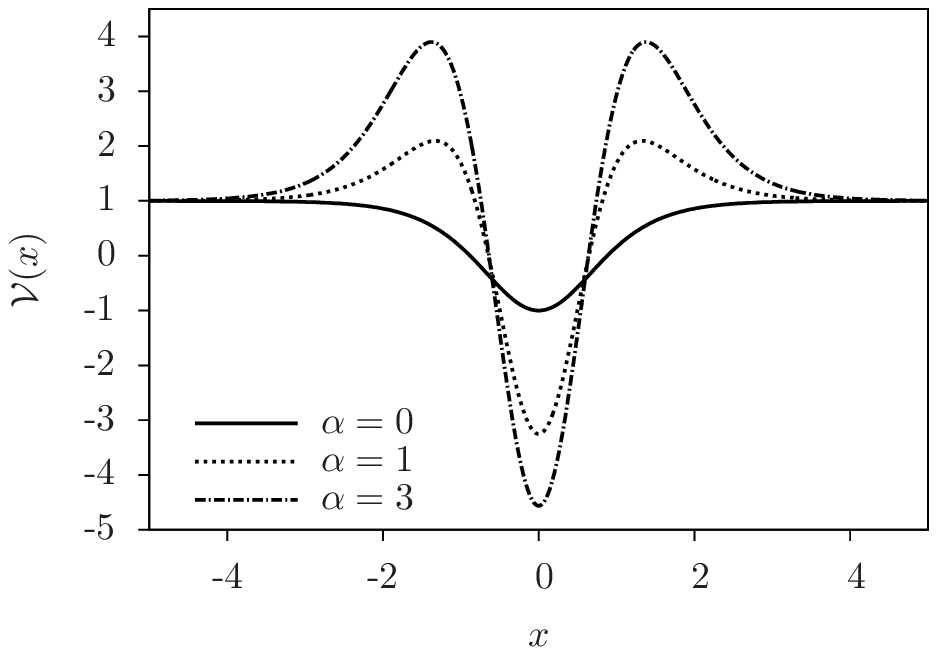}
\includegraphics[width=0.48\textwidth]{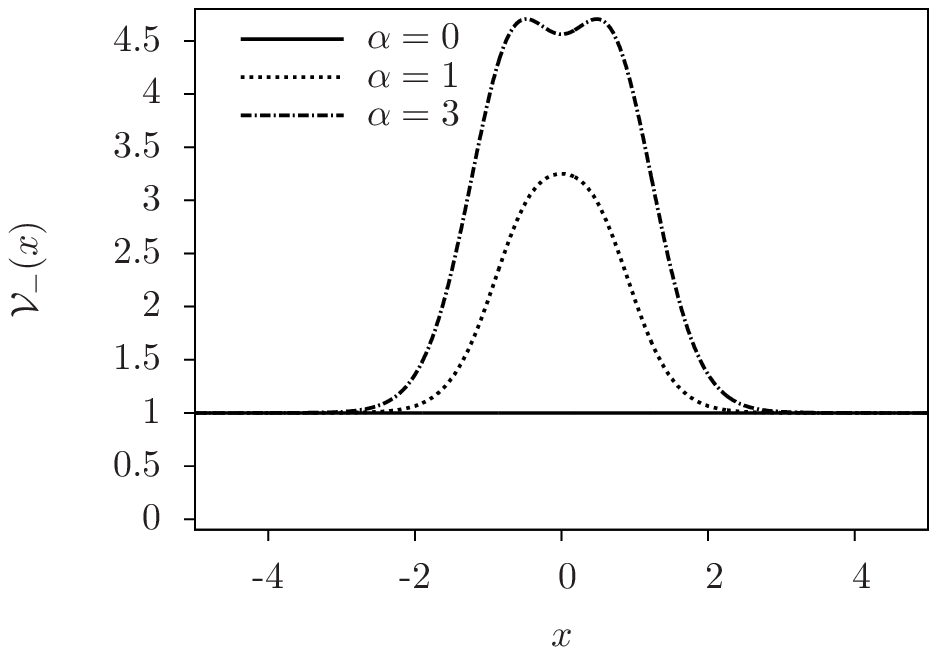}
\end{center}
\caption{Plots of $\mathcal{V}(x)$ and $\mathcal{V}_{-}(x)$ for the generalized Sine-Gordon kink solution of the $X^2$ model ($k=1$ and $v=1.2$).}
\label{figureSGveff}
\end{figure*}

The shape of the effective potential $\mathcal{V}(x)$\footnote{We consider $\mathcal{V}(x)$ rather than $\mathcal{V}(x^\ast)$ because we can always obtain the analytical expression for the former. Besides, the shapes of the two potentials are almost the same: they are the same potential defined on two different coordinate systems.} is plotted in figure~\ref{figureSGveff}. We see that when $\alpha=0$, the potential is a P\"oschl-Teller type potential (see ref.~\cite{CooperKhareSukhatme1995} for more details), while for $\alpha>0$ the potential becomes a volcano one.
It is worth to mention that in many brane world models, the volcano potential plays an important role as the mechanism for localizing gravity. An interesting property of such potential is the possibility for finding massive resonant modes~\cite{Gremm2000}. However, the existence of resonant modes cannot be directly judged from the shape of $\mathcal{V}$. For instance, in refs.~\cite{Gremm2000,KehagiasTamvakis2001}, although the graviton is attracted by volcano potentials, no graviton resonance was found.

According to the theory of supersymmetric quantum mechanics, if there are resonant modes in the spectrum of $H$, there should also be the resonant modes of the partner Hamiltonian $H_{-}$. That means, if there is a resonant mode, both $\mathcal{V}$ and $\mathcal{V}_{-}$ should have double barrier structure. It is easier for us to judge the existence of resonances from the shape of $\mathcal{V}_{-}$, (for example see ref.~\cite{ZhongLiuZhao2014}).

To be explicity, we draw the $\mathcal{V}_{-}$ corresponding to the generalized Sine-Gordon kink.
When $\alpha=0$, we see $\mathcal{V}_{-}(x)=k^2$ is a constant. So in this case the mass spectrum is constituted by the zero mode and a continuum of states (the mass gap is $k^2$). All the states with $\omega_n>k^2$ are plane waves.

However, as one introduces the $X^2$ term, $\mathcal{V}_-$ deforms from a trivial line firstly to a lump for a relatively small $\alpha$ ($\alpha < 1$). Then, as $\alpha$ increases a double barrier structure appears, which is a sign for massive resonant modes.

\subsection{The generalized $Z_2$ kink}
\begin{figure*}
\begin{center}
\includegraphics[width=0.48\textwidth]{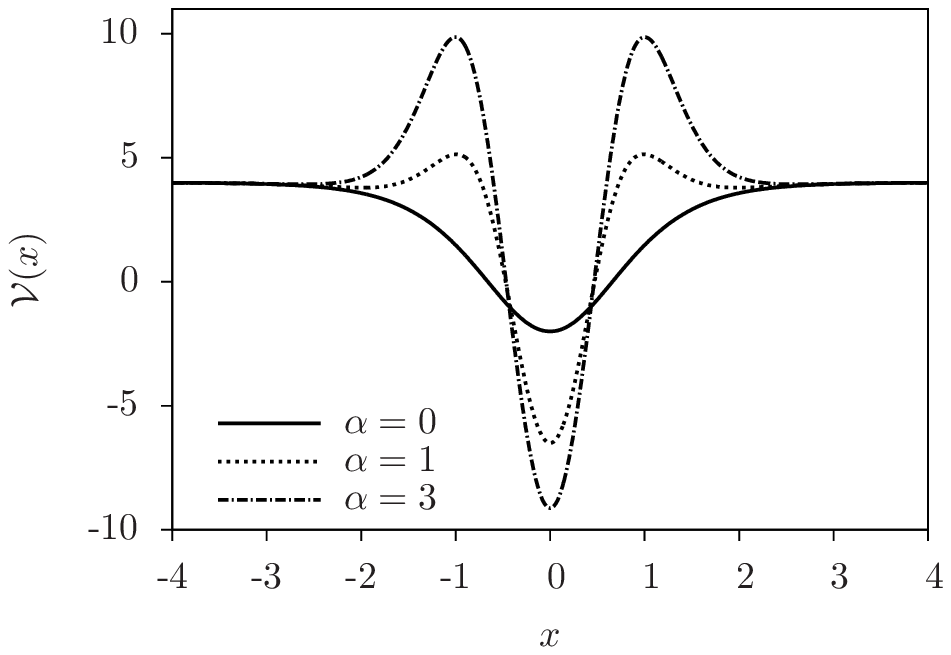}
\includegraphics[width=0.48\textwidth]{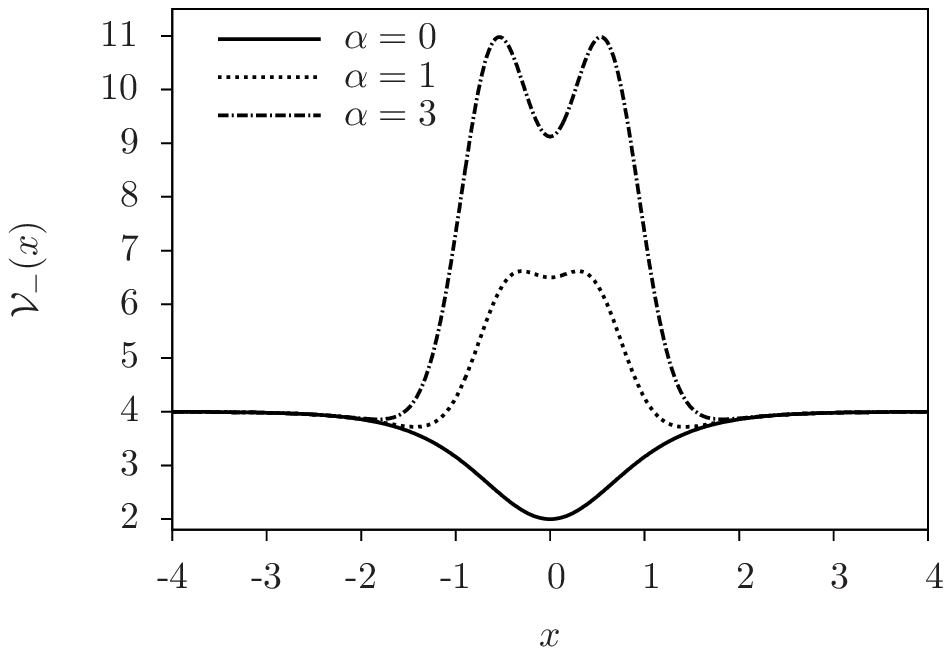}
\end{center}
\caption{Plots of $\mathcal{V}(x)$ and $\mathcal{V}_{-}(x)$ for the generalized $Z_2$ kink solution of the $X^2$ model ($k=1$ and $v=1.2$).}
\label{figurePhi4veff}
\end{figure*}

Now let us turn to the $Z_2$ kink. The effective potential $\mathcal{V}(x)$ and the corresponding partner potential $\mathcal{V}_{-}(x)$ are depicted in figure~\ref{figurePhi4veff}. Same to the generalized Sine-Gordon kink, when $\alpha$ increases, the potential $\mathcal{V}(x)$ of the $Z_2$ kink deforms from the original P\"oschol-Teller potential to volcano potentials. In addition, double barrier structure emerges in $\mathcal{V}_{-}(x)$ for large $\alpha$.

As depicted in the right panel of figure~\ref{figurePhi4veff}, when $\alpha=0$, $\mathcal{V}_{-}(x)$ is a P\"oschl-Teller type potential. This consists with our knowledge that $H_{-}$ should have one massive bound state. However, as $\alpha$ increases, the potential well of $\mathcal{V}_{-}(x)$ gradually deforms into one or a few barriers. This means that the original massive bound state disappears for large $\alpha$. On the other hand, for a larger $\alpha$, one probably would find massive resonant modes. The study on resonances needs numerical calculation, so we will not carry it out in this paper.

\section{Summary}
In this paper, we studied the solitary wave solutions of a class of two-dimensional noncanonical scalar field models. We first derived the equation for the linear perturbation around an arbitrary solitary wave solution of the system. We found that the equation derived from the quadratic action is equivalent with the one derived directly from the equation of motion. This equivalence breaks down if gravity is turned on. Then we constructed first-order formalisms for two typical types of noncanonical models. With these formalisms, one can easily construct exactly solvable models as well as the corresponding solitary wave solutions. We considered the $X^2$ model as a toy model, and constructed two solvable models: the generalized Sine-Gordon model and the generalized $Z_2$ model. In addition to the analytical solutions, we also investigated the linear structures of these kink solutions. We found that when the $X^2$ term is introduced, the perturbation modes feel a volcano potential rather than the original P\"oschl-Teller potential. As a result, massive resonant modes might be produced in these generalized models. Besides, the massive bound state in the standard $Z_2$ model disappears in a model with large $X^2$ term.

\section*{Acknowledgments}

This work was supported by the Program for New Century Excellent Talents in University, the National Natural Science Foundation of China (Grants No. 11075065, No. 11375075), and the Fundamental Research Funds for the Central Universities (Grant No. lzujbky-2013-18). Y.Z. was also supported by the scholarship granted by the Chinese Scholarship Council (CSC).


\providecommand{\href}[2]{#2}\begingroup\raggedright\endgroup
\end{document}